\documentclass[twocolumn,showpacs,preprintnumbers,amsmath,amssymb]{revtex4}

\usepackage{graphicx}
\usepackage{dcolumn}
\usepackage{bm}
\usepackage{graphics}
\usepackage{amsmath}
\usepackage{amssymb}
\usepackage{amscd}
\usepackage{afterpage}
\usepackage{float,times}
\usepackage{subfigure}
\usepackage{rotating}
\usepackage{multirow}
\usepackage{fancyheadings}
\usepackage{epsfig}
\usepackage{theorem}
\usepackage{moreverb}
\usepackage{euscript}
\usepackage{psfrag}

\begin{document}

\title{(3+2) Neutrino Scheme From A Singular Double See-Saw Mechanism}

\author{K. L. McDonald}\email{k.mcdonald@physics.unimelb.edu.au}

\author{B. H. J. McKellar}
 \email{b.mckellar@physics.unimelb.edu.au}
\author{A. Mastrano}%
 \email{amastran@physics.unimelb.edu.au}
\affiliation{%
School of Physics, Research Centre for High Energy Physics, The
University of Melbourne, Victoria, 3010, Australia\\
}%

\date{\today}

\begin{abstract}

We obtain a $3+2$ neutrino spectrum within a left-right symmetric
framework by invoking a singular double see-saw mechanism. Higgs
doublets are employed to break $SU_{R}(2)$ and three additional
fermions, singlets under the left-right symmetric gauge group, are
included. The introduction of a singularity into the singlet fermion
Majorana mass matrix results in a light neutrino sector of three
neutrinos containing predominantly $\nu_{\alpha L}$,
$\alpha=e,\mu,\tau$, separated from two neutrinos containing a small
$\nu_{\alpha L}$ component. The resulting active-sterile mixing in the
$5\times 5$ mixing matrix is specified once the mass eigenvalues and
the $3\times3$ submatrix corresponding to the PMNS mixing matrix are
known.

\end{abstract}

\pacs{14.60.Pq, 14.60.St}
\maketitle

\section{\label{sec:intro}INTRODUCTION\protect\\}
Our understanding of neutrino masses and mixings has rapidly improved
in recent years, with solar~\cite{solar},
atmospheric~\cite{atmospheric} and terrestrial~\cite{kamland,k2k,
  chooz,palo_verde} neutrino oscillation experiments providing
valuable insight. The reactor experiments
CHOOZ~\cite{chooz} and Palo Verde~\cite{palo_verde} indicate that
the atmospheric and solar oscillations are effectively
decoupled~\cite{bilenky_giunti_solar_atm_decoupled} and the totality of
the data suggests that the atmospheric and solar anomalies can be
adequately explained by
three flavour neutrino mixing. The reported
$\bar{\nu}_{\mu}-\bar{\nu}_{e}$ oscillation signal of LSND~\cite{lsnd}
provides an interesting piece of oscillation data that conflicts with this three
flavour explanation. The ongoing MiniBooNE~\cite{miniboone} experiment
will soon test the LSND result.   

The $3+1$ and $2+2$ neutrino spectra arose from a minimalistic approach to
the simultaneous resolution of the solar,
atmospheric and LSND
neutrino data in
terms of neutrino oscillations. The neutrino
spectrum is extended in a minimal fashion via the addition of one sterile neutrino
state. Currently favoured fits to the solar and
atmospheric data in terms of purely active neutrino oscillations
leave little room for additional sterile
states~\cite{maltoni_schwetz_tortola_valle}. Recent high precision
measurements of the S-factor (defined
in~\cite{s_factor_defined}) by the Seattle
group~\cite{seattle_group_solar_boron_flux} give $S_{17}(0)=22.1\pm
0.6$~eV-b, leading to an expected $^{8}B$ solar neutrino flux 13\%
larger than that measured by SNO~\cite{sno_solar_result_used_by_gai}
(for a discussion see~\cite{Gai:2004da,Davids:2003aw}). Questions
regarding the distinction between atmospheric $\nu_{\mu}\rightarrow
\nu_{\tau}$ and $\nu_{\mu}\rightarrow \nu_{s}$ transitions~\cite{foot_superk_multiring}
 and the absolute statistical significance of some data
 fits~\cite{foot_ruling_out_four_neutrino_models} have also been
 raised. The $3+1$ and
$2+2$ schemes come into conflict with the data in different ways. The
source of incompatibility for $2+2$ spectra comes from relations amongst the sterile
components in the atmospheric and solar neutrinos that are difficult to
reconcile with experimental results (it has been suggested that global
fits to data that include the effects of small mixing angles, usually
neglected in analysis, are required to invalidate the $2+2$
schemes~\cite{pas_song_weiler}). $3+1$ spectra, on the other hand, are disfavoured by
comparisons of short-baseline disappearance
data~\cite{cdhs_sbl,bugey_sbl} with the LSND result.

The study of $3+2$ spectra follows the minimalistic attitude that
motivated the four neutrino models and data fits. The addition of the
second sterile state can simultaneously enhance the predicted LSND
signal and relax the laboratory and atmospheric bounds on the mixing
matrix elements $U_{e4}$ and $U_{\mu
  4}$~\cite{peres_smirnov_three_plus_two}. The second sterile state is
required to mix with both $\nu_{e}$ and $\nu_{\mu}$ to contribute to
the LSND probability and avoid opening up new channels for $\nu_{e}$
or $\nu_{\mu}$ disappearance. Provided $m_{5}^{2}>\Delta m^{2}_{LSND}$
the bounds on $U_{e4}$ and $U_{\mu 4}$ are modified and the second
$\Delta m^{2}$ will contribute to LSND. The splitting between the two
predominantly sterile states $\Delta m^{2}_{45}$ should also be
resolved by LSND to ensure the LSND signal is enhanced. If, for
example, $\Delta m^{2}_{14}\sim(1-2)~\mathrm{eV}^{2}$ and $\Delta
m^{2}_{15}>8~\mathrm{eV}^{2}$, the LSND signal can be enhanced whilst
relaxing the short-baseline
constraints~\cite{peres_smirnov_three_plus_two}.
The statistical analysis of~\cite{sorel_conrad_shaevitz} suggests that
if $\Delta m^{2}_{15}\sim 22~\mathrm{eV}^{2}$ the predicted LSND
signal may be enhanced by $\mbox{60-70\%}$. Using horizontal
symmetries, $3+2$ spectra with see-saw suppressed light sterile
neutrinos have been studied~\cite{Babu:2003is}, whilst the coexistence
of large active-active and large active-sterile mixing in $3+2$
scenarios was studied in~\cite{mckellar_stephenson_goldman_garbutt}.

Though minimalistic, the
introduction of two sterile states seems counterintuitive to the
suggestive demands of a familial quark-lepton symmetry. The latter
makes the addition of three
right-handed neutrinos to the standard model seem a logical
extension. The discovery of a quark-lepton familial symmetry may hint
at an underlying left-right symmetric gauge theory, broken to the
standard model at some high energy scale. In this note our objective
is to theoretically motivate a $3+2$ neutrino model within a
left-right symmetric framework. Higgs doublets are employed, rather
than triplets, to break $SU_{R}(2)$ and additional singlet neutral
fermions, sterile under the gauge symmetries, are included. The $3+2$
spectrum results from the introduction of a singularity into the
singlet fermion Majorana mass matrix. The resulting modified double
see-saw mechanism produces a light neutrino sector with three
predominantly $SU_{L}(2)$ active neutrinos separated from two
neutrinos predominantly sterile under $SU_{L}(2)$.

The structure of this note is as follows. In Section \ref{sec:extending_sm} the particle content of the model is presented in conjunction with a brief discussion of the double see-saw mechanism. The neutrino content of the model receives focus in Section \ref{sec:singular_double_seesaw} where the eigenstates are derived.  Section \ref{singular_double_seesaw_scales} contains a discussion of the scales required to make the resulting neutrino spectrum experimentally feasible and some concluding remarks can be found in Section \ref{sec:discussion}.
%-------------------------------------------------------------------------------%
\section{Extending the Standard Model\label{sec:extending_sm}}
The left-right symmetric model, with gauge group
$\mathcal{G}_{LR}=SU_{C}(3)\times SU_{L}(2)\times SU_{R}(2)\times
U_{B-L}(1)$, is considered a natural extension of the standard model
(SM). The addition of three right-handed neutrinos to the SM fermionic
spectrum automatically qualifies $SU_{R}(2)$ as a gaugeable
symmetry. The use of Higgs triplets to break $SU_{R}(2)$ at a high
energy scale provides a theoretical framework for the realisation of
the see-saw mechanism~\cite{see_saw}, coupling the existence of light
predominantly $SU_{L}(2)$ doublet neutrinos to the existence of (as
yet undetected) heavy gauge
bosons~\cite{mohapatra_senjanovic_left_right_with_triplets}. An
alternative path to massive neutrinos within a left-right symmetric
framework relies on Higgs doublets, rather than triplets, to achieve
the high energy breaking of
$SU_{R}(2)$~\cite{mohapatra_left_right_with_doublets,left_right_with_zero_left_doublet_vev}.
This path denies the see-saw mechanism the opportunity to explain the
lightness of the $SU_{L}(2)$ active neutrinos, but the inclusion of
extra neutral fermions, singlets under $\mathcal{G}_{LR}$,  can
provide an alternative explanation.
The addition of three such singlets leads to a leptonic Yukawa lagrangian of the form:

\begin{eqnarray}
{\mathcal{L}}_Y=h_{ij}^{1}{\bar{L}}_L^i\phi L_R^j + h_{ij}^{2}\bar{L}_L^{i}\tilde{\phi}
L_R^{j}+ M_{Sij} \bar{S^{i}}^c S^{j}\nonumber\\
+f_{ij}(\bar{L}^{ci}_L X_L S^{j} + \bar{L}^{ci}_R X_R S^{j})
+h.c.,\label{3_2_LR_model_lagrangian}
\end{eqnarray}
where $L_{L,R}$ are the fermion doublets, $S$ denotes the singlet
fermions, $\phi$ is a Higgs bidoublet and $X_{L,R}$ are Higgs doublets, ie:
\[L_L \sim (1,2,1,-1), L_R \sim (1,1,2,-1),\]
\[S \sim (1,1,1,0),\]
\[\phi \sim (1,2,2,0),\]
\[X_L \sim (1,2,1,1), X_R \sim (1,1,2,1).\]
Denoting the second Pauli matrix by $\tau_{2}$ the bidoublet $\tilde{\phi}$ in (\ref{3_2_LR_model_lagrangian}) is given by $\tilde{\phi}=\tau_{2}\phi^{*}\tau_{2}$. The singlet neutrinos have bare Majorana mass terms, whilst the
doublet neutrinos acquire Dirac mass couplings to the singlets only if
$X_{L,R}$ develop non-zero VEV's. One requires a non-zero value for
$\langle X_R\rangle$ to break $SU_{R}(2)$ at some high scale, but may
take $\langle X_L\rangle
=0$~\cite{left_right_with_zero_left_doublet_vev} to preclude Dirac
mass terms coupling
$\nu_{L}^{i}$ to the singlets. In the basis $(\nu_L, \nu_R^c, S^c)$
the neutral fermion mass matrix has the
form:

\begin{eqnarray}
\left(\begin{array}{ccc} 0 & m_{LR} & 0 \\m_{LR}^T &
0 & M_{RS}\\0 & M_{RS}^T & M_{S}
\end{array}\right),\label{double_see_saw_mass_matrix}
\end{eqnarray}
where $m_{LR}$ and $M_{RS}$ are Dirac  mass matrices and $M_{S}$ is
the singlet Majorana mass matrix. We shall denote the scale of
non-zero entries in $m_{LR}$, $M_{RS}$ and $M_{S}$ as $m$, $M$ and
$\mu$ respectively. The Dirac mass matrix $m_{LR}$ ($M_{RS}$) arises
when $\phi$ ($X_{R}$) acquires a VEV. The physical condition $\langle
X_{R}\rangle\gg\langle\phi\rangle$ implies $M\gg m$, though the
relationship between $M$ and $\mu$ is not predetermined. The effective
light neutrino mass matrix is given by
$M_\nu=-m_{LR}^T(M_{RS}^{-1})^T M_{S}(M_{RS}^{-1})m_{LR}$. Some interesting scale hierarchies are:\\
$\bullet \mu \ll M$\\
This provides a further suppressing factor of $\frac{\mu}{M}$ to the
light neutrino mass scale relative to that of the see-saw mechanism,
$\sim\frac{m^{2}}{M}$. Consequently the suppressing scale $M$ can be
set lower. This case has been referred to as an inverse see-saw mechanism (see for example~\cite{Deppisch:2004fa}). The small value required of $\mu$ in this case is considered natural in the technical sense~\cite{thooft_cargese_lectures} as lepton number conservation is restored in the limit $\mu\rightarrow 0$.\\
$\bullet\mu \gg M$\\
This hierarchy generates a see-saw mechanism between the $\nu_{R}$'s
and the $S$'s, giving an effective right-handed neutrino scale of
order $\frac{M^2}{\mu}$.\\
We note that neutrino mass matrices of this form are found in some
string inspired models~\cite{mohapatra_valle_double_seesaw} and
provide a basis for the so called double see-saw
mechanism~\cite{king_review,smirnov_neutrino_review}.
%-------------------------------------------------------------------%
\section{Singular Double See-Saw Mechanism\label{sec:singular_double_seesaw}}
In this paper we investigate the double see-saw mechanism, with the
hierarchy $\mu\gg M$, when the singlet Majorana mass matrix $M_{S}$ is
of rank 2. The Dirac mass matrices $m$ and $M$ will remain general. The $9\times 9$ mass matrix
(\ref{double_see_saw_mass_matrix}) will then lead to 9 Majorana
neutrinos as follows:\\
$\bullet$ Two ultra-heavy Majorana neutrinos of order $\mu$,
predominantly containing the fully sterile singlets,\\
$\bullet$ A pseudo-Dirac pair of heavy neutrinos of order $M$. These
will be an admixture of the right handed neutrinos and the massless singlet,\\
$\bullet$ A lighter pair of Majorana neutrinos with mass $\sim
M^{2}/\mu$, containing mostly $\nu_{R}$'s with a small $\nu_{L}$
component,\\
$\bullet$ Three Majorana neutrinos of order $m^{2}/(M^{2}/\mu)$. These
will be mostly $\nu_{L}$'s with a small $\nu_{R}$ component.\\

We begin by performing a singular see-saw analysis~\cite{allen_john_etc,chun_kim_lee,McDonald:2004qx} on the submatrix:
\begin{eqnarray}
\left(\begin{array}{cc}
0 & M_{RS} \\M_{RS}^T & M_{S}
\end{array}\right).
\end{eqnarray}
The $9\times 9$ mass matrix is repartitioned and the submatrix $M_{S}$
diagonalised as follows:
\[M_{\nu(9\times 9)}= \left(\begin{array}{ccc} 0 & m_{LR} & 0 \\m_{LR}^T &
0 & M_{RS}\\0 & M_{RS}^T & M_{S}
\end{array}\right)\equiv \left(\begin{array}{cc} A_{6\times 6} &
  \beta_{6\times 3} \\(\beta_{6\times 3})^{T} & M_{S}
\end{array}\right)\]
\[=\left(\begin{array}{cc} I_{6\times 6} & 0_{6\times 3} \\0_{3\times 6} & R_1^T
\end{array}\right)\left(\begin{array}{cc} A & \beta R_1^T \\R\beta^T &
  M_{S}^{diag} \end{array}\right)\left(\begin{array}{cc} I_{6\times
    6} & 0_{6\times 3} \\0_{3\times 6} & R_1
\end{array}\right),\]
where $M_{S}^{diag}$ has the zero eigenvalue of $M_{S}$ in the (1,1)
entry. A further repartition to separate the zero eigenvalue of $M_{S}$
gives:
\[
M_\nu =\left(\begin{array}{cc} A & \beta R_1^T \\R\beta^T &
M_{S}^{diag} \end{array}\right)=\left(\begin{array}{cc}
A'_{7\times7} & B_{7\times2} \\(B_{7\times2})^{T} & \omega_{2\times2}
\end{array}\right),\]
where $A'$ has the zero eigenvalue of $M_{S}$ in its lower right corner. Next, we block diagonalise $M_\nu$:
\begin{eqnarray}
M_\nu &=&\left(\begin{array}{cc}
A'_{7\times7} & B_{7\times2} \\B^T_{2\times7} & \omega_{2\times2}
\end{array}\right)\nonumber\\
&=&S\left(\begin{array}{cc}
Q_{7\times7} & 0_{7\times2} \\0_{2\times7} & \omega_{2\times2}
\end{array}\right)S^{T},\label{block_diagonalized_7by7_2by2}
\end{eqnarray}
where:
\[S=\left(\begin{array}{cc}
I_{7\times7} & P_{1(7\times2)} \\-(P_{1(7\times2)})^{T} & I_{2\times2}
\end{array}\right),\]
and:
\[P_1=B\omega^{-1},\]
\begin{eqnarray}Q=A'-P_1\omega P_1^T=A'-B\omega^{-1}B^{T}.\label{decompostion_of_Q}\end{eqnarray}
Equation (\ref{block_diagonalized_7by7_2by2}) demonstrates that to
order $M^{2}/\mu$ the eigenvalues of $M_\nu$ include two heavy
Majorana neutrinos, with masses of order $\mu$, which are linear
combinations of the sterile states $S^{i}$. The second term in the
above expression for $Q$ provides a
see-saw type correction to the submatrix $A'$. The matrix $Q$ has the form:
\[Q=\left(\begin{array}{cc} 0_{3\times3} & \gamma_{3\times4}
    \\(\gamma_{3\times4})^{T} & \omega_{2(4\times4)}  \end{array}\right),\]
where the non-zero elements of $\gamma$ are of order $m$ and $\omega_{2}$
contains non-zero elements of order $M$ and order $M^{2}/\mu\ll
M$. The matrix $\omega_{2}$ must now be diagonalised. A perturbative treatment gives
the zeroth order eigenvalues as
$\omega_{2}^{diag}=(0,0,\lambda^{(0)}_{\omega_{2}},-\lambda^{(0)}_{\omega_{2}})$,
where $\lambda^{0}_{\omega_{2}}\sim M$. The eigenvectors corresponding
to the zero eigenvalues are linear combinations of the $\nu_{R}$'s,
whilst the Dirac pair contains the orthogonal
combinations of $\nu_{R}$'s and the zero eigenvector of $M_{S}$. The
correction term $-B\omega^{-1}B^{T}$ of equation
(\ref{decompostion_of_Q}) splits the degenerate non-zero eigenvalues,
forming a pseudo-Dirac pair, and gives a mass of order $M^{2}/\mu$ to
the zero eigenvalues. The diagonalisation of $\omega_{2}$ gives:
\[Q=\left(\begin{array}{cc} 0_{3\times3} & \gamma_{3\times4}
    \\\gamma^T_{4\times3} & \omega_{2(4\times4)}  \end{array}\right)\]
\[=\left(\begin{array}{cc} I_{3\times3} & 0_{3\times4} \\0_{4\times3}
    & R^T_{2(4\times4)}  \end{array}\right)\left(\begin{array}{cc}
    0_{3\times3} & \gamma R_2^T \\R_2\gamma^T & \omega_{2}^{diag}
  \end{array}\right)\left(\begin{array}{cc} I_{3\times3} &
    0_{3\times4} \\0_{4\times3} & R_{2(4\times4)}
  \end{array}\right),\] 
and we repartition the above matrix to:
\begin{eqnarray}
\left(\begin{array}{cc} 0_{3\times3} & \gamma R^T \\R\gamma^T & \omega_{2}^{diag}
\end{array}\right)=\left(\begin{array}{cc} \Omega_{5\times5} & B_2 \\B_2^T & \omega'_{2}
\end{array}\right),\label{partioned_to_5by5}
\end{eqnarray} 
where $\omega'_{2}$ contains the two eigenvalues of order $M$ from
$\omega_{2}^{diag}$ and the order $M^{2}/\mu$ eigenvalues form the lower
right block of $\Omega$. Finally we block diagonalise (\ref{partioned_to_5by5}):
\begin{eqnarray}
\left(\begin{array}{cc} \Omega_{5\times5} & B_2 \\B_2^T & \omega'_{2}
\end{array}\right)=\mathcal{A}\left(\begin{array}{cc} \Omega'_{5\times5} & 0 \\0 & \omega'_{2}
\end{array}\right)\mathcal{A}^{T}
\end{eqnarray}
where:
\begin{eqnarray}
\mathcal{A}=\left(\begin{array}{cc} I_{5\times5} & P_{2(5\times2)} \\-P_{2(2\times5)}^T &
I_{2\times2}
\end{array}\right),
\end{eqnarray}
and:
\[P_2=B_2 {\omega'_2}^{-1},\]
\begin{eqnarray}
\Omega'=\Omega-P_2 {\omega'_2}^{-1}P^T_2=\Omega-B_2 {\omega'_2}^{-1}
B^T_2.\label{decomposition_of_omega_prime}
\end{eqnarray}
$\Omega'$ has three eigenvalues of order $m^{2}/(M^{2}/\mu)$ and
two of order $M^{2}/\mu$. Contributions from the term $-B_2 {\omega'_2}^{-1}
B^T_2$ in (\ref{decomposition_of_omega_prime}) are negligible as they are
of order $m^{2}/M$. Thus $\Omega'\approx \Omega$ and the light sector
mass matrix has the form:
\begin{eqnarray}
M_{Light}&=&\Omega'\approx\Omega\nonumber\\
&=&\left( \begin{array}{cc}0_{3\times3} & \tilde{m}_{3\times 2}\\
    (\tilde{m}_{3\times 2})^{T} & \tilde{M}\end{array}\right),\label{defining_m_light}
\end{eqnarray}
in the basis $(\nu_{e},\nu_{\mu},\nu_{\tau},\nu_{1R}'^{c},\nu_{2R}'^{c})$. In
(\ref{defining_m_light}) the matrix $\tilde{m}$ contains rotated entries of the
Dirac mass matrix
$m_{LR}$ and $\tilde{M}=\mathrm{diag}(\lambda_{1},\lambda_{2})$ with
the lighter eigenvalues of $\omega_{2}^{diag}$ denoted as
$\lambda_{1,2}\sim M^{2}/\mu$. Block diagonalising $M_{Light}$ gives:
\begin{eqnarray}
M_{Light}&=&\mathcal{R}\left( \begin{array}{cc} \tilde{m}_{3\times
      2}\tilde{M}^{-1}(\tilde{m}_{3\times 2})^{T} & 0\\ 0&
      \tilde{M}\end{array}\right)\mathcal{R}^{T},\label{block_diagonalizing_m_light}
\end{eqnarray}
with:
\begin{eqnarray}
\mathcal{R}=\left(\begin{array}{cc} I_{3\times3} & \tilde{P}_{3\times2} \\-(\tilde{P}_{3\times2})^T &
I_{2\times2}
\end{array}\right),
\end{eqnarray}
and
$\tilde{P}=\tilde{m}\tilde{M}^{-1}$.
Eq. (\ref{block_diagonalizing_m_light}) demonstrates that
$\lambda_{1,2}\sim M^{2}/\mu$ are approximate eigenvalues of $M_{\nu(9\times9)}$ with
eigenvectors containing mostly $\nu_{1,2R}'^{c}$ and a small
$\nu_{\alpha L}$, $\alpha=e,\mu,\tau$, component. The remaining three eigenvalues are
found by diagonalising $\tilde{m}_{3\times2}\tilde{M}^{-1}(\tilde{m}_{3\times2})^{T}$ and are thus $\sim
    m^{2}/(M^{2}/\mu)$. The eigenvectors are predominantly composed of
    the states $\nu_{\alpha L}$. Mixing between the states
    $\nu_{\alpha L}$ and $\nu_{1,2R}'^{c}$ is controlled by
    $\tilde{P}$ and is of order $m/(M^{2}/\mu)$. 
%-------------------------------------------------------------------------------------%
\section{Singular Double See-Saw Mechanism Scales\label{singular_double_seesaw_scales}}
The five lightest mass eigenstates generated by the singular
double see-saw mechanism are structured in a suitable manner to
realise a $3+2$ neutrino spectrum. The scales required to make this $3+2$
spectrum experimentally feasible are now discussed.

LSND requires mass-squared differences of $\sim1-10~\mathrm{eV}^{2}$. In the
above model the relevant scales for LSND are set by
$\lambda_{1,2}\sim M^{2}/\mu$. Consequently $M^{2}/\mu\sim
1$~eV is required and the LSND result alone permits a freedom to scale
both $M$ and $\mu$. As $M\sim\langle X_{R}\rangle$ is related to the mass of the gauge bosons
coupled to the right-handed currents it is experimentally constrained. Considerations of the $K_{L}-K_{S}$ mass difference have given
the bound $M_{W_{R}}\ge 1.6$~TeV~\cite{Beall:1981ze}, which implies
$M\gtrsim 1.6$~TeV, assuming that the coupling constants are not unreasonably small. This in turn gives $\mu\ge10^{15}$~GeV. Note that if $M\sim1$~TeV then
$\mu\sim10^{15}$~GeV which hints that the $S_{i}$'s may acquire mass at
a GUT scale. The bounds on $M$ and $\mu$ decouple the eigenstates with
masses of order $M$ and $\mu$ from the low energy
phenomenology.

The predominantly active states have a mass of order
$m^{2}/(M^{2}/\mu)$. The atmospheric neutrino bound of
$\sqrt{\Delta m^{2}_{atm}}\approx 5\times10^{-2}$~eV together with the relationship
$M^2/\mu\sim 1$~eV implies $m\gtrsim 10^{-1}$~eV. With
$m\sim10^{-1}$~eV and $M^{2}/\mu\sim 1$~eV the active-sterile mixing
from $\tilde{P}$ is of order $10^{-1}$.

To fit the atmospheric and solar oscillation data in terms of active
flavour oscillations requires:
\begin{eqnarray}
\tilde{m}_{3\times2}\tilde{M}^{-1}(\tilde{m}_{3\times2})^{T}=UM^{diag}U^{T},
\end{eqnarray}
where the $3\times 3$ mixing matrix $U$ has the approximate form of
the experimentally measured $U_{PMNS}$ matrix. The full $5\times 5$
mixing matrix is:
\begin{eqnarray}
U_{5\times5}=\left(\begin{array}{cc} I_{3\times3} & \tilde{P}_{3\times2} \\-(\tilde{P}_{3\times2})^T &
I_{2\times2}
\end{array}\right)\left(\begin{array}{cc} U & 0 \\0 &
I_{2\times2}
\end{array}\right).
\end{eqnarray}
Previous works suggest that taking $\Delta m^{2}_{14}\sim
1~\mathrm{eV}^{2}$ and $\Delta m^{2}_{15}> 8~\mathrm{eV}^{2}$ can
enhance the LSND signal and alleviate the mixing matrix element bounds
applicable to $3+1$ models from the other short base-line
experiments~\cite{peres_smirnov_three_plus_two,sorel_conrad_shaevitz}.
To demonstrate the type of active-sterile mixing matrix elements
obtained in this model we have numerically determined them for some
mass values. Taking $\theta_{12}=\pi/6$, $\theta_{13}=0$ and
$\theta_{23}=\pi/4$ for the PMNS angles we took the predominantly
active mass eigenvalues to be:
\begin{eqnarray}
& &M^{diag}\nonumber\\
&=&U^{T}\tilde{m}_{3\times2}\tilde{M}^{-1}(\tilde{m}_{3\times2})^{T}U\nonumber\\
&=&\mathrm{diag}(0,0.014,0.045),\label{diagonalizing_light_active_matrix}
\end{eqnarray}
where the masses are in eV. The diagonalisation of
(\ref{diagonalizing_light_active_matrix}) is enforced with $\Delta
m_{41}^2=0.92$ eV$^2$ and $\Delta m_{51}^2$
set at the best fit value obtained in~\cite{sorel_conrad_shaevitz}, $\Delta m_{51}^2=22$
eV$^2$. The resulting five neutrino mixing matrix is:

\begin{eqnarray}U_{\alpha i}=\left(\begin{array}{ccccc}  0.866&0.500&0&-0.094&0.032 \\
-0.359&0.612&0.707&-0.292&0.145\\0.359&-0.612&0.707&-0.062&0.067\\0&0.188&0.250&1&0\\0&-0.064&-0.150&0&1
\end{array}\right)\label{singular_double_see_saw_mixing_numbers}\end{eqnarray}\\
These values are to be compared with the best fit values $U_{e4}=0.121$,
$U_{e5}=0.036$, $U_{\mu 4}=0.204$ and $U_{\mu 5}=0.224$ obtained
in~\cite{sorel_conrad_shaevitz}. Both the electron and muon neutrinos couple more strongly to
the fourth mass eigenstate in
(\ref{singular_double_see_saw_mixing_numbers}), though reducing
$M_{5}$ increases the elements $|U_{e,\mu 5}|$, as expected given the
form of $\tilde{P}$. The greatest deviation from the values in~\cite{sorel_conrad_shaevitz} occurs for the elements $U_{\mu4,5}$. When calculating the probability $P(\bar{\nu}_{\mu}\rightarrow\bar{\nu}_{e})$ relevant for LSND the mixing matrix elements always occur in the combinations $U_{e4}U_{\mu 4}$ and $U_{e5}U_{\mu 5}$. The values in (\ref{singular_double_see_saw_mixing_numbers}) give $|U_{e4}U_{\mu 4}|=2.7\times10^{-2}$ and $|U_{e5}U_{\mu 5}|=4.6\times10^{-3}$ whilst the values in~\cite{sorel_conrad_shaevitz} give $2.4\times10^{-2}$ and $8.1\times10^{-3}$ respectively. The smaller value $|U_{e4}|=0.094<0.121$ is seen to compensate somewhat for the deviation of $|U_{\mu 4}|=0.292>0.204$. With the values in (\ref{singular_double_see_saw_mixing_numbers}) the contribution of the heavier sterile state to the LSND signal is lower than that obtained with the best fit values in~\cite{sorel_conrad_shaevitz}. The contribution is  still large enough however to enhance significantly the LSND signal relative to a $3+1$ oscillation pattern. The above numbers are presented as an example
only and the compatibility of this model with the best fit values in~\cite{sorel_conrad_shaevitz} depends on the size of the light mass eigenvalues. Deviations from the values $\theta_{12}=\pi/6$, $\theta_{13}=0$ and
$\theta_{23}=\pi/4$ also shift the active-sterile mixing but the dependence of this mixing on the size of the light eigenvalues is generally stronger. We emphasise that within this framework, knowledge of the
mass eigenvalues and the experimentally extractable elements of the
$3\times3$ PMNS matrix U specifies the size of the active-sterile
mixing.
%--------------------------------------------------------------------------%
\section{Discussion\label{sec:discussion}}
From a cosmological point of view this model, along with all models
that attempt to explain LSND by the addition of sterile neutrinos,
risks thermalising the sterile states and disrupting the standard
BBN. We adopt the view that an adjustment of the standard paradigm
will be required if the oscillation interpretation of the LSND result
is confirmed, alleviating the need to comply with the derived bounds
of $N_{\nu}<4$ from the $^{4}$He
abundance~\cite{walker_steigman_schramm_olive_kang,cyburt_fields_olive} and $\sum
m_{\nu}<0.7-1.0$~eV from cosmological data~\cite{mass_sum_limit}.

We have not commented on the origin of the singularity in the singlet
Majorana mass matrix. Many methods of obtaining singularities in mass
matrices exist in the literature (for
example~\cite{partially_broken_ssm,babu_mohapatra_glashow_model,davidson}).
Some methods which may be relevant for application to our model
include a horizontal gauge symmetry in the singlet
sector~\cite{foot_king} and the supersymmetric realisation of a
singular Majorana mass matrix of Du and Liu~\cite{du_liu}. The method advocated in~\cite{Grimus:2004hf} to generate arbitrary mass matrix texture zeros could also be employed.

We have shown that the singular double see-saw mechanism can be
employed to generate a $3+2$ neutrino spectrum within a left-right
symmetric framework. The resulting active-sterile mixing is found to
be determined once the mass eigenvalues and the submatrix of the
$5\times5$ mixing matrix corresponding to the PMNS matrix are
specified. From a model building point of view this result may be of
interest if MiniBooNe confirms the oscillation interpretation of the
LSND result.
 
%-------------------------------------------------------------------------------%
\section*{Acknowledgements}
K.M. thanks Robert Foot for useful discussions. This work was
supported in part by the Australian Research Council.
%---------------------------------------------------------------------%

%-----------------------------------------------------------------------------%
\end{document}